\documentclass{aastex7}

\definecolor{black}{rgb}{0,0,0}
\definecolor{red}{rgb}{1.0,0,0}
\newcommand\T{\rule{0pt}{3.1ex}}       % Top strut
\newcommand\B{\rule[-1.7ex]{0pt}{0pt}} % Bottom strut

\newcommand{\omm}{1I/`Oumuamua}

\newcommand{\BLBerkeley}{Breakthrough Listen, University of California, Berkeley, 501 Campbell Hall \#3411, Berkeley, CA 94720, USA}
\newcommand{\UCB}{Department of Astronomy, University of California, Berkeley, 501 Campbell Hall \#3411, Berkeley, CA 94720, USA}
\newcommand{\seti}{SETI Institute, 339 Bernardo Ave, Suite 200, Mountain View, CA 94043, USA}
\newcommand{\KZA}{University of Malta, Institute of Space Sciences and Astronomy}

\newcommand{\BLOxford}{Breakthrough Listen, University of Oxford, Denys Wilkinson Building, Keble Road, Oxford OX1 3RH, UK}
\newcommand{\Manchester}{Department of Physics and Astronomy, University of Manchester, UK}
\newcommand{\Marshall}{Department of Mathematics and Physics, Marshall University, Huntington, WV 25755, USA}

\begin{document}

\title{Breakthrough Listen Observations of 3I/ATLAS with the Green Bank Telescope at 1--12 GHz}

%% Note that the corresponding author command and emails has to come
%% before everything else. Also place all the emails in the \email
%% command instead of using multiple \email calls.

% Create Author List
\correspondingauthor{Ben Jacobson-Bell}
\email{benjacobsonbell@berkeley.edu}

\author[0009-0009-6231-9280]{Ben Jacobson-Bell}
\email{benjacobsonbell@berkeley.edu}
\affiliation{\UCB}
\affiliation{\BLBerkeley}
%\affiliation{\BLOxford}

\author[0000-0003-4823-129X]{Steve Croft}
\email{scroft@berkeley.edu}
\affiliation{\BLBerkeley}
\affiliation{\BLOxford}
\affiliation{\seti}

\author[0000-0002-9112-1734]{Ellie White}
\email{elliewhite1420@gmail.com}
\affiliation{\Marshall}

\author[0000-0003-2828-7720]{Andrew P. V. Siemion}
\email{siemion@berkeley.edu}
\affiliation{\UCB}
\affiliation{\BLOxford}
\affiliation{\seti}
\affiliation{\KZA}
\affiliation{\Manchester}

\author[0000-0002-7042-7566]{Matthew Lebofsky}
\email{lebofsky@berkeley.edu}
\affiliation{\UCB}
\affiliation{\BLBerkeley}

\author[0000-0001-6950-5072]{David H. E. MacMahon}
\email{davidm@astro.berkeley.edu}
\affiliation{\UCB}
\affiliation{\BLBerkeley}


\begin{abstract}
    3I/ATLAS, an interstellar object, made its closest approach to Earth on 2025 December 19. On 2025 December 18, the Breakthrough Listen program conducted a technosignature search toward 3I/ATLAS using the 100\,m Robert C. Byrd Green Bank Telescope at 1--12\,GHz. We report a nondetection of candidate signals down to the 100\,mW level.
\end{abstract}

\section{Introduction}
\label{sec:intro}

3I/ATLAS (also designated C/2025 N1 (ATLAS) and, formerly, A11pl3Z) is the third interstellar object (ISO), following \omm\ and 2I/Borisov, to be discovered during a passage through the Solar System. The Asteroid Terrestrial-impact Last Alert System (ATLAS) reported its discovery on 2025 July 1 \citep{deen2025comet}. 

Unlike \omm, 3I/ATLAS exhibits mostly typical cometary characteristics \citep{deen2025comet}, including a coma and an unelongated nucleus. There is currently no evidence to suggest that ISOs are anything other than natural astrophysical objects. However, given the small number of such objects known (only three to date), and the plausibility of interstellar probes as a technosignature \citep[e.g,][]{Freitas85}, thorough study is warranted \citep[see][]{Davenport25}. Putative nonanthropogenic interstellar probes are likely to communicate via narrowband radio signals for transmission efficiency and for the low extinction of such signals across interstellar space; all of humanity's spacecraft, including the now-interstellar craft \textit{Voyager 1} and \textit{Voyager 2}, communicate via such signals.

The Breakthrough Listen (BL) program observed 3I/ATLAS using the 100-m Robert C.\ Byrd Green Bank Telescope (GBT) on UT 2025 December 18, $\sim$1 day before the ISO's closest approach to Earth. Similar technosignature searches have recently been undertaken by \citet{ATA:ATLAS} and \citet{Pisano25} over different frequency ranges and with different sensitivities.
Like those searches, we find no credible detections of narrowband radio technosignatures originating from 3I/ATLAS.

\section{Observations}
\label{sec:obs}

Our GBT observations cover four bands of the radio spectrum, each with a different receiver: $L$ (1.1--1.9\,GHz), $S$ (1.8--2.7\,GHz), $C$ (4.0--7.8\,GHz) and $X$ (7.6--11.7\,GHz).  With each receiver, we observed a 30-min cadence: three 5-min on-target pointings interspersed with three 5-min off-target pointings in an ABACAD arrangement. All observations were conducted between UT 04:15 and 09:15 on 2025 December 18.

We used the BL back end \citep{MacMahon18, Lebofsky19} to digitize the data and the software \verb+turboSETI+ \citep{turboSETI} to conduct our signal search. We refer the reader to \citet{Lebofsky19} for an overview of our observing strategy and to \cite{Enriquez17} and \cite{Choza_2024} for further details on \verb+turboSETI+.

\begin{figure*}
    \centering
    \includegraphics[width=0.795\textwidth]{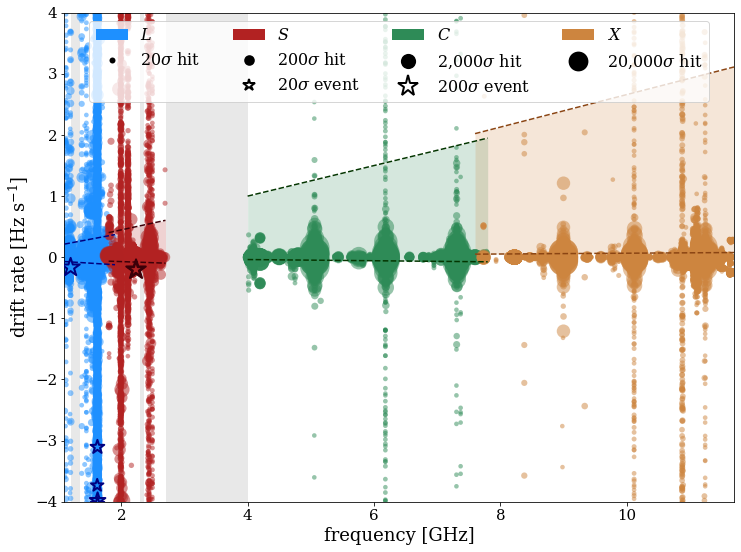}
    \caption{Distribution of hits (circles) and events (stars) over frequency and drift rate. The marker size gives the S/N. The gray-shaded regions show ranges not sampled, including narrow notch filter regions in the $L$ and $S$ bands. The color-shaded regions give the range of drift rates expected from Earth's orbital motion and rotational motion and 3I/ATLAS's rotation at each of the four observing bands. These regions do not perfectly align between bands due to 3I/ATLAS's radial acceleration changing during overhead time between observations. No events lie in the expected drift rate regions.}
    \label{hit_event_dist}
\end{figure*}

\section{Results}
\label{sec:conclusion}

The BL back end channelizes our observation data to 2.8\,Hz resolution. Using \verb+turboSETI+, we search for signals above a $\sim$16$\sigma$ threshold \citep[corresponding to a software input threshold of 5$\sigma$; see][]{Choza_2024} in a drift rate range of $\pm$4\,Hz\,s$^{-1}$. We use a lower threshold, 10$\sigma$ (software input 3$\sigma$), for off-target scans to discourage the ``detection'' of candidates with slightly less power in those scans. We find 471,198 ``hits,'' or drifting signals above 16$\sigma$ (for on-target scans) or 10$\sigma$ (for off-target scans). After applying the sky localization filter on the full ABACAD cadences, which removes all candidates that appear in one or more off-target scans, we are left with nine ``events.''

Figure \ref{hit_event_dist} shows the distribution of hits and events over frequency and drift rate, with marker size encoding signal-to-noise ratio (S/N). The shaded regions give the range of drift rates expected due to Earth's rotational and orbital motion and 3I/ATLAS's rotation \citep{ATLAS-rotation}, assuming an unmodulated transmission. Overdensities in frequency correspond to known bands of high radio-frequency interference (RFI) contamination. We visually inspect the nine events and, due to their appearance in the off-target scans and/or congruence to known contaminants, rule out all of them as RFI.

We place new constraints on the existence of radio transmitters at the location of 3I/ATLAS. 
The minimum flux density detectable using our survey, assuming an unresolved signal, is \citep{Gajjar_2021_GC}
\begin{equation}
    S_{\text{min}} = \frac{(\text{S/N})_0}{\beta} \frac{\text{SEFD}}{\delta\nu_t} \sqrt{\frac{\delta\nu}{{n_{\text{pol}}}\,\tau_{\text{obs}}}},
\end{equation}
where $(\text{S/N})_0$ is the S/N threshold of the survey (16); SEFD is the system equivalent flux density of the receiver (10--15\,Jy); $\delta\nu_t$ is the transmitted signal width (assumed to be $\sim$1\,Hz); $\delta\nu$ is the survey frequency resolution (2.8\,Hz); $n_{\text{pol}}$ is the number of polarizations (2); $\tau_{\text{obs}}$ is the length of an observation (300\,s); and $\beta$ is a dedrifting efficiency parameter that accounts for loss due to signal smearing at high drift rates. The effective isotropic radiated power (EIRP) corresponding to a flux density $S$ is
\begin{equation}
    \text{EIRP} = 4\pi d^2 S \,\delta\nu_t,
\end{equation}
where $d$ is the distance of the transmitter ($\sim$1.8\,AU for 3I/ATLAS during our observations). Our survey concludes that there are no isotropic continuous-wave transmitters above 0.1\,W at the location of 3I/ATLAS. For comparison, a cell phone is an approximately isotropic continuous-wave transmitter at a level of $\sim$1\,W.

All data collected by BL will be publicly available. The observations described here can be downloaded from the \href{https://bldata.berkeley.edu/ATLAS}{Breakthrough Listen Data Portal}.

\begin{acknowledgements}
    Funding for BL is provided by the \href{http://breakthroughinitiatives.org}{Breakthrough Prize Foundation}. 
    E.{}W.\ is supported by the National Science Foundation
Graduate Research Fellowship Program under Grant
No.~1000383199. Any opinions, findings, and conclusions or recommendations expressed in this material are those of the author(s) and do not necessarily reflect the views of the National Science Foundation.
\end{acknowledgements}

\bibliographystyle{aasjournal}
\bibliography{main}

\end{document}